\documentclass[%
 reprint,
 superscriptaddress,
 amsmath,amssymb,
 aps,
floatfix,
]{revtex4-2}

\usepackage[hidelinks]{hyperref}
\hypersetup{
    colorlinks=true, 
    linktoc=all,     
    linkcolor=blue, 
    citecolor=blue, 
    urlcolor=black,
}
\usepackage{graphicx}  
\usepackage{dcolumn}   
\usepackage{bm}        
\usepackage{amssymb}   
\usepackage{subcaption}
\usepackage{ragged2e}
\usepackage{amsmath}
\usepackage{xcolor}
\usepackage{orcidlink}
\usepackage[export]{adjustbox}
\usepackage{float}
\restylefloat{table}
\usepackage{xr}

\begin{document}



\title{
Frictional contact of soft polymeric shells
}%

\author{Riad Sahli~\orcidlink{0000-0001-9077-3602}}%
\affiliation{%
Department of Mechanical and Production Engineering, Aarhus University, Denmark}%
\author{Jeppe Mikkelsen}%
\affiliation{Department of Mechanical and Production Engineering, Aarhus University, Denmark}%
\author{Mathias Sätherström Boye}%
\affiliation{Department of Mechanical and Production Engineering, Aarhus University, Denmark}%
\author{Marcelo A. Dias~\orcidlink{0000-0002-1668-0501}}
\email[Corresponding Author: ]{Marcelo.Dias@ed.ac.uk}
\affiliation{Institute for Infrastructure and Environment, School of Engineering, The University of Edinburgh, EH9 3FG Edinburgh, UK}
\author{Ramin Aghababaei~\orcidlink{0000-0002-0700-0084}}%
\email[Corresponding Author: ]{ra@mpe.au.dk}
\affiliation{%
Department of Mechanical and Production Engineering, Aarhus University, Denmark}%

                         
\begin{abstract} The classical Hertzian contact model establishes a monotonic correlation between contact force and area. Here, we showed that the interplay between local friction and structural instability can deliberately lead to unconventional contact behavior when a soft elastic shell comes into contact with a flat surface. The deviation from Hertzian contact first arises from bending within the contact area, followed by the second transition induced by buckling, resulting in a notable decrease in the contact area despite increased contact force. Friction delays both transitions and introduces hysteresis during unloading. However, a high amount of friction suppresses both buckling and dissipation. Different contact regimes are discussed in terms of rolling and sliding mechanisms, providing insights for tailoring contact behaviors in soft shells.

\end{abstract}



\maketitle

\begin{figure}[b]
    \centering
    \begin{tabular}{cc}
     \adjustbox{valign=b}{\begin{tabular}{@{}c@{}}
    {%
    \textbf{a)}
         \includegraphics[page=1,width=.4\linewidth]{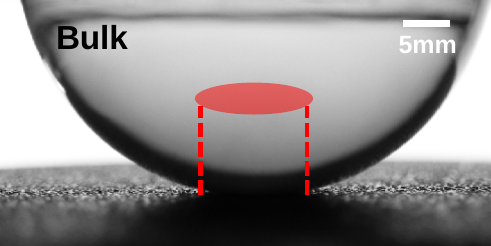}} \\
         \textbf{b)}
    {%
         \includegraphics[page=2,width=.4\linewidth]{All_case_07.pdf}}
    \end{tabular}}
     &
     \adjustbox{valign=b}{{%
     \textbf{c)}   \includegraphics[page=1,width=.45\linewidth,height=3.7cm]{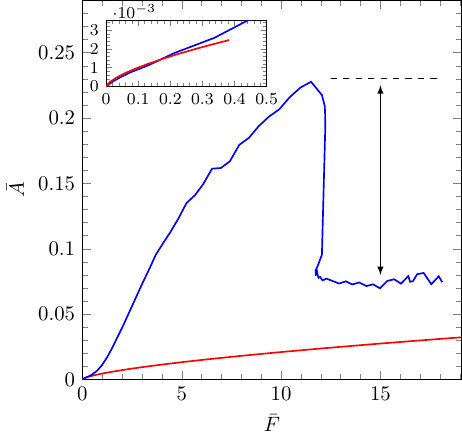}}}
    
    \end{tabular}
        \caption[justification=justified]{(a) Bulk hemisphere made of PDMS in contact with a PLA plate. The contact area is circular and grows with normal load with an exponent of 2/3, following Hertzian contact solution (red line in subfig.~c). (b) PDMS hemispherical shell in dry contact with a PLA plate (solid blue line in subfig.~c.). (c) Presents the variation of contact area ($\bar{A}=2 A/(\pi R^2)$) as a function of normal load ($\bar{F}=F R/(E h^{3})$) for bulk and shell. First, a smooth deviation from Hertzian contact can be observed (inset).  As indentation progresses, the contact morphology transitions from circular to annular, causing a sudden drop in the contact area (marked by a double arrow).
\justifying }
    \label{fig:Fig0}
  \end{figure}

The contact area and pressure distribution between solid surfaces directly influence the frictional~\cite{bowden2001friction,archard1957elastic} and thermal resistance~\cite{madhusudana1996thermal} as well as the sealing capacity~\cite{persson2004nature} of the interface. Assuming a parabolic pressure profile, infinitesimal deformation, and material homogeneity, Hertzian theory~\cite{hertz1882ueber} establishes a power-law relationship between the normal load and contact area for a wide range of elastic geometries, including spherical, cylindrical, and elliptical contacts~\cite{popov2019handbook} (also see Fig.~\ref{fig:Fig0}). Alternatively, it is shown that the contact area increases linearly with the normal load when plastic deformation occurs ~\cite{bowden2001friction}. Since these findings, the contact area-load relation faced a transformative modification as the interplay of other factors such as friction ~\cite{CARLSSON2000,STORAKERS2005, Weber2022}, adhesion~\cite{pastewka2014contact,bico2020, Oliver2023}, roughness ~\cite{greenwood1967,luan2005breakdown, Aghababaei2022, Muser2022}, and material and geometric nonlinearities~\cite{weber2018molecular, WU201671, Ghaednia2017} were examined. It is well established that the contact area grows with the normal load \cite{dieterich1994direct}, while material properties, contact geometry, and surface conditions dictate the growth rate ~\cite{sahli2018evolution,sahli2019shear, MATSUDA2018, Mo2009}.

Recent studies~\cite{Slesarenko2024, Aymard2024} have discussed the possibilities of modifying friction laws (i.e. the correlation between the normal and frictional forces) by designing metainterfaces with certain roughness morphology. Studying the contact response of soft polymeric shells has offered another angle to modify the contact laws using the structural instability at the contact. Pauchard and Rica demonstrated a sub-critical bifurcation, namely a first-order phase transition, changes the nature of the mechanical response in contact~\cite{pauchard1998contact}. It was observed that a flat contact exists initially between the rigid plane and the shell. However, it transitions to a post-buckling state characterized by contact along an axisymmetric circular ridge formed between the rigid plane and the inverted shell. The relevant parameters of this problem are material, shell's Young's modulus $E$, and geometrical, shell's radius $R$ and thickness $h$. The combination of these results into the relevant force scale, $F_{0}\equiv Eh^3/R$, and a detailed analysis proposes three regimes~\cite{audoly2010elasticity}: (i) Hertzian regime with a circilar contact area, for a range of forces $F\ll F_{0}$, where deformation occurs within a region smaller than the shell thickness, $h$; (ii) intermediate regime, where $F\sim F_{0}$ and equilibrium is obtained by a balance between bending and stretching, the shell flattens in a manner assumed to be similar to Hertzian contact with a disk-like contact area; (iii) post-buckling regimes, where $F>F_{0}$ and the stretching energy caused by sphere flattening becomes significant, destabilizing the shell and buckling into an inverted shape, which is akin to the isometric problem presented by Pogorelov~\cite{pogorelov1988bendings}. Recent studies ~\cite{VAZIRI2009, C3SM50279A, knoche2014secondary} have demonstrated secondary buckling instability, where a transition from axisymmetric to asymmetric deformation mode with multiple vertices occurs. Additionally, the effect of structural imperfections on the buckling mode has also been investigated ~\cite{Lee2016}. Despite these advances in understanding the mechanics of shell indentation, there still exist limitations in the exploration of the effect of structural instability and friction on contact morphology and pressure distribution.

In this letter, we examine the evolution of contact morphology and pressure when a thin shell comes into contact with a rigid, flat surface at different stages of indentation with axisymmetric deformation. Our experiments and numerical simulations show that a circular contact area is first established and grows by indentation, followed by a smooth transition from circular to disk-like contact as a result of bending in the contact area. As indentation progresses, a secondary transition occurs, resulting in a sharp reduction in the contact area despite increased contact force. Our results led to three novel discoveries: (1) the force-contact area deviates from Hertzian solution in the intermediate regime, where the contact geometry transitions from circular to disk-like contact; (2) a non-monotonic force-contact area relation can be achieved in connection with the shell instability (post-buckling regime), which is dictated by the local friction; and (3) friction introduces hysteresis upon unloading with a maximum at a critical coefficient of friction (CoF), after which hysteresis drops as high friction suppresses sliding. One critical aspect of our findings, which pertains to the contact behavior in the intermediate regime is that we observed a departure from previous claims that full contact is maintained until the bifurcation point~\cite{audoly2010elasticity}. Additionally, it is also shown that local friction delays and eventually suppresses the post-bifurcation regime, hindering the abrupt change in the force-contact area relation and hysteresis.
%
\begin{figure}[t]
     \centering
         \includegraphics[width=1\linewidth]{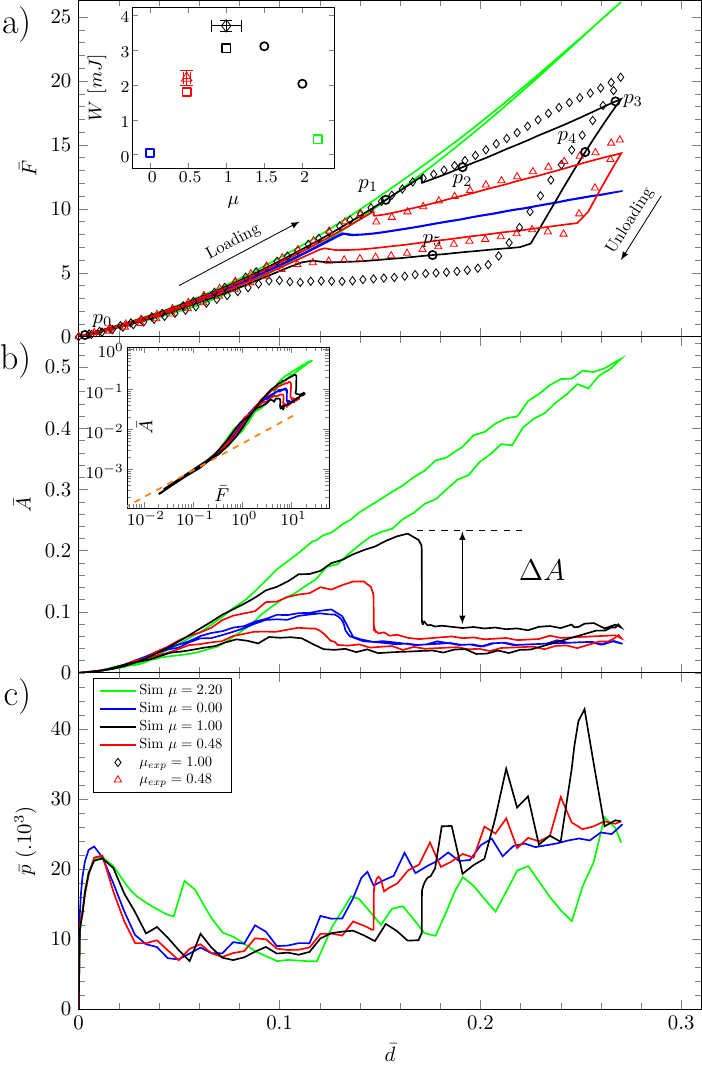}
     \caption[justification=justified]{The evolution of (a) normalized force $\bar{F}$, (b) contact area $\bar{A}$, and (c) maximum contact pressure multiplied by $10^3$ for clarity, ($\bar{p} =P_{max}/E$) as a function of indentation ($\bar{d}=d/R$) for different CoF. Black diamond and red triangle symbols in (a) present, respectively, the experimental data under dry and lubricated conditions. Solid lines represent simulation results for $\mu =0$ (blue), $0.48$ (red), $1.0$ (black), and $2.2$ (green). The inset in (a) presents the dissipated energy during unloading $W$, highlighting a parabolic correlation with CoF. The inset in (b) reveals two transitional points, 1) the deviation of force-area relation from the Hertzian solution (the orange dash line) due to bending, and 2) a sudden drop in contact area due to shell buckling.\justifying}
     \label{fig:Fig1}
\end{figure}

The snapshots in Fig.~\ref{fig:Fig0}a and b showcase the deformation modes of a bulk solid and a shell, both made from cross-linked polydimethylsiloxane (PDMS), upon contact with a flat polylactic acid (PLA) plate. Fig.~\ref{fig:Fig0}c presents the force-contact area relation, highlighting an early deviation from Hertzian contact, followed by a sudden drop in the contact area when a transition occurs in the deformation mode.
First, we experimentally study the contact of a thin, elastic hemispherical shell made of cross-linked PDMS~\cite{le2009comparison} with a flat PLA plate, having a roughness of $Sa = 15$ $\mu m$ to minimize the adhesion effect (Fig.~\ref{fig:Fig0}). The PDMS shell, with a radius of $R=25 mm$ and thickness $h = 1.33 \pm 0.06$ mm, is made by pouring PDMS mixture over a $25$ mm radius steel ball~\cite{lee2016fabrication}. In a displacement-control manner (see supplementary Fig.~\ref{figSI:fullsetup}
\cite{SuppInf}), the PLA plate comes into contact with the shell at a constant velocity of $10$ mm/min, where the force-displacement data are recorded. A side camera is used to monitor the deformation of the shell. To study the effect of friction coefficient, we ran the test under dry and lubricated conditions, where ``Kema GL-68"  is used as the lubricant. Additional experiments are conducted to measure $\mu$ between PDMS and PLA under dry ($1.0 \pm 0.2$) and lubricated ($0.48 \pm 0.09$) contacts (see supplementary Fig.~\ref{figSI:friction}~\cite{SuppInf}). 

\begin{figure*}[t]
    \begin{subfigure}[!]{0.03\textwidth}
    \centering
    \textbf{a)}
  \end{subfigure}
     \centering
     \begin{subfigure}[b]{0.156\textwidth}
         \centering
         \includegraphics[page=1,width=\textwidth]{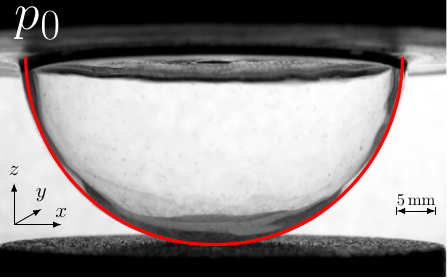}
     \end{subfigure}
     \hfill
          \begin{subfigure}[b]{0.156\textwidth}
         \centering
         \includegraphics[page=2,width=\textwidth]{Frame_preparation_Version3.pdf}
     \end{subfigure}
     \hfill
          \begin{subfigure}[b]{0.156\textwidth}
         \centering
         \includegraphics[page=3,width=\textwidth]{Frame_preparation_Version3.pdf}
     \end{subfigure}
          \hfill
          \begin{subfigure}[b]{0.156\textwidth}
         \centering
         \includegraphics[page=4,width=\textwidth]{Frame_preparation_Version3.pdf}
     \end{subfigure}
     \hfill
          \begin{subfigure}[b]{0.156\textwidth}
         \centering
         \includegraphics[page=5,width=\textwidth]{Frame_preparation_Version3.pdf}
     \end{subfigure}
          \hfill
          \begin{subfigure}[b]{0.156\textwidth}
         \centering
         \includegraphics[page=6,width=\textwidth]{Frame_preparation_Version3.pdf}
     \end{subfigure}
     
  \begin{subfigure}[t]{0.03\textwidth}
    \textbf{b)}
  \end{subfigure}
    \begin{subfigure}[b]{0.156\textwidth}
         \centering
         \includegraphics[page=1,width=\textwidth]{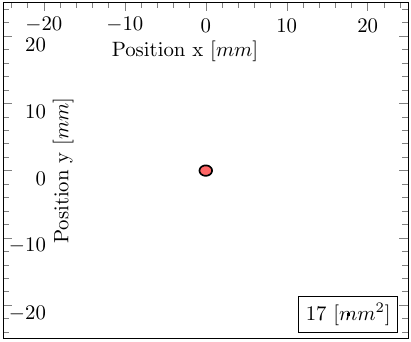}
     \end{subfigure}
     \hfill
          \begin{subfigure}[b]{0.156\textwidth}
         \centering
         \includegraphics[page=2,width=\textwidth]{plot_circle_Version04.pdf}
     \end{subfigure}
     \hfill
        \begin{subfigure}[b]{0.156\textwidth}
         \centering
         \includegraphics[page=3,width=\textwidth]{plot_circle_Version04.pdf}
     \end{subfigure}
          \hfill
          \begin{subfigure}[b]{0.156\textwidth}
         \centering
         \includegraphics[page=4,width=\textwidth]{plot_circle_Version04.pdf}
     \end{subfigure}
          \hfill
        \begin{subfigure}[b]{0.156\textwidth}
         \centering
         \includegraphics[page=5,width=\textwidth]{plot_circle_Version04.pdf}
     \end{subfigure}
     \hfill
        \begin{subfigure}[b]{0.156\textwidth}
         \centering
         \includegraphics[page=6,width=\textwidth]{plot_circle_Version04.pdf}
     \end{subfigure}

  \begin{subfigure}[t]{0.03\textwidth}
    \textbf{c)~}
  \end{subfigure}
     \begin{subfigure}[b]{0.156\textwidth}
         \centering
         \includegraphics[page=1,width=\textwidth]{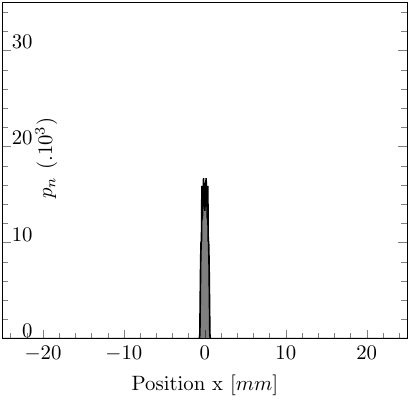}
     \end{subfigure}
     \hfill
          \begin{subfigure}[b]{0.156\textwidth}
         \centering
         \includegraphics[page=2,width=\textwidth]{plot_pression_Version04.pdf}
     \end{subfigure}
     \hfill
        \begin{subfigure}[b]{0.156\textwidth}
         \centering
         \includegraphics[page=3,width=\textwidth]{plot_pression_Version04.pdf}
     \end{subfigure}
          \hfill
          \begin{subfigure}[b]{0.156\textwidth}
         \centering
         \includegraphics[page=4,width=\textwidth]{plot_pression_Version04.pdf}
     \end{subfigure}
     \hfill
        \begin{subfigure}[b]{0.156\textwidth}
         \centering
         \includegraphics[page=5,width=\textwidth]{plot_pression_Version04.pdf}
     \end{subfigure}
          \hfill
        \begin{subfigure}[b]{0.156\textwidth}
         \centering
         \includegraphics[page=6,width=\textwidth]{plot_pression_Version04.pdf}
     \end{subfigure}

    \caption[justification=justified]{(a) In-situ snapshots of shell profile indented by a PLA plate, under dry contact ($\mu=1.0$) at different stages of indentation, corresponding to points $p_0$-$p_5$, marked in Fig.~ \ref{fig:Fig1}a. The profiles obtained from the simulation are superimposed with red solid lines for comparison. (b) and (c) present, respectively, the corresponding contact morphology and the contact pressure distribution ($p(x)/E$), obtained from simulations multiplied by $10^3$ for clarity. Inset numbers in (b) present the total contact area. The evolution of contact area and pressure distribution, as well as the indentation force and shell profile for different CoFs can be seen in supplementary movies~\cite{SuppInf}).\justifying}
    \label{fig:Fig2}
\end{figure*}
To complement our experiments with details of contact area and pressure evolution over a wider range of $\mu$, we performed systematic finite element simulations using the commercial software ``Abaqus"\cite{abaqus2021abaqus}. The numerical setup, shown in supplementary Fig.~\ref{figSI:abacus}~\cite{SuppInf}, simulated the contact between an axisymmetric PDMS shell with a rigid plate. Details of loading and boundary conditions as well as material properties are presented in the Supplementary section~\cite{SuppInf}. 


Fig.~\ref{fig:Fig1}a summarizes the experimental and numerical results, presenting the evolution of contact force ($\bar{F}=F R/(E h^{3})$) during the loading and unloading stages under dry ($\mu=1.0$) and lubricated ($\mu=0.48$) conditions. We additionally include simulation results for the frictionless ($\mu=0.0$) and high-friction ($\mu=2.2$) scenarios. Despite the absence of the adhesion effect in our simulation, a close agreement between the experimental and numerical results has been obtained for the dry and lubricated contacts in all stages of indentation. As seen, the contact force monotonically increases with the indentation until buckling occurs, after which a small drop is identified in the indentation force. The force continues to increase at a lower rate in the post-buckling regime. One can see that the evolution of force in the post-buckling regime is a function of CoF. As friction rises, it opposes the compressive stresses within the shell, increasing the critical force needed for instability initiation. In other words, the heightened frictional forces hinder relative motion between surfaces, strengthening resistance to compressive loading. Consequently, the critical force threshold rises due to the combined effects of friction-induced resistance and the reaction to compressive stresses on the shell material. This observation highlights how frictional mechanics control the structural stability of materials.

A significant dissipation through hysteresis is also observed upon unloading in both cases. It is evident that as CoF increases, the buckling is delayed, while a larger energy dissipation occurs upon unloading. As expected, no dissipation occurs in the frictionless case, indicating that sliding in the post-buckling stage is the main source of energy dissipation. Interestingly, it can be seen that high CoF can completely avoid shell buckling and, consequently, energy dissipation is reduced to a minimum. The inset of Fig.~\ref{fig:Fig1}a show a parabolic relation between the dissipated energy ($W = \oint_S F dx$) and CoF, indicating the existence of a critical CoF, where sliding and dissipation in the post-buckling stage are maximized (also see Table.\ref{tab:1}~\cite{SuppInf}). This critical CoF may be a function of geometrical and material parameters. 

The evolution of contact area in different stages of indentation is presented in Fig.~\ref{fig:Fig1}b. It can be seen that the normalized contact area ($\bar{A}=2 A/(\pi R^2)$) increases monotonically until the point of instability, after which a sharp reduction occurs (also see Table.\ref{tab:1}~\cite{SuppInf}. For the case of high friction ($\mu=2.2$) and the absence of instability, the contact area increases for the entire simulation. The inset figure in Fig.~\ref{fig:Fig1}b, compares the force-contact area relationship with the Hertzian solution shown by an orange dashline in the loading regime. Interestingly, it can be seen that deviation from the Hertzian solution occurs in the intermediate regime and before the point of buckling. This observation challenges the previous assumption of Hertzian-like behavior in the intermediate regime \cite{audoly2010elasticity}. Remarkably, the contact area remains constant in the post-buckling regime despite the increase in the indentation force. This stands in contrast to the contact response of a bulk sphere, where the contact area monotonically grows ~\cite{hertz1882ueber,sahli2018evolution} (also see Fig.~\ref{fig:Fig0}). Three regimes of deformation can also be distinguished by monitoring the average contact pressure as a function of indentation. Fig.~\ref{fig:Fig1}c presents the maximum contact pressure normalized by elastic modulus ($\bar{p} =P_{avg}/E$). The pressure initially increases in the Hertzian regime (regime I), followed by a smooth drop and saturation (regime II). Clearly, it can be seen that the pressure in the intermediate regime II does not follow the Hertzian behavior shown in regime I. The maximum pressure experiences a sharp increase as a result of buckling and the reduction in the contact area (see Fig.~\ref{fig:Fig1}b), indicating the transition into the post-buckling regime III.

An in-situ analysis of shell deformation and the evolution of contact area and pressure for the dry contact scenario ($\mu = 1.0$), corresponding to marked points $p_0$-$p_5$ in Fig.~\ref{fig:Fig1}a, is illustrated in Fig.~\ref{fig:Fig2} (see supplementary movies~\cite{SuppInf} for other CoFs). It can be seen that the numerical simulations predict the profile of the shell in all regimes. At the onset of contact ($p_0$), the circular contact geometry (Fig.~\ref{fig:Fig2}b) and the parabolic contact pressure distribution (Fig.~\ref{fig:Fig2}c) follow the Hertzian solution. $p_1$ demonstrates the deviation from the Hertzian solution in the intermediate regime but before the point of buckling by showing a disk-like contact area and non-parabolic pressure distribution as a direct result of bending in the shell. A similar pressure profile has also been derived theoretically ~\cite{updike1972pressure}. Contrary to the parabolic Hertzian pressure distribution, the pressure in the intermediate regime reduces from the leading edge to the trailing edge (see Supp. Movies~\cite{SuppInf}
), indicating that the change in the contact morphology is dominated by the rolling mechanism. This explains the minimization of dissipation until the point of buckling, after which sliding and thus dissipation occur at the contact. In the post-buckling regime ($p_2$ and $p_3$), the contact occurs along a circular ridge with a parabolic pressure distribution across the contact width. In the unloading phase, the outer side of the shell is first unloaded elastically ($p_4$) as shown in Supp. Fig.~\ref{figSI:Fig2}. In this regime, while the buckled (inner) side of the shell remains fairly stationary due to friction (see Supp. Fig.~\ref{figSI:radius}a), the outer side unrolls, causing a reduction in the contact area (Fig.~\ref{figSI:radius}b). The unloading phase continues with a transition, where the local tangential load overcomes friction. In this regime ($p_5$), while the contact area remains constant (see Fig.~\ref{fig:Fig1}b), dissipation occurs via frictional sliding. Eventually, the shell unbuckles, after which the Hertzian solution is recovered. 

\begin{figure}[t]
    \begin{subfigure}[b]{\linewidth}
         \centering
         \includegraphics[width=\linewidth]{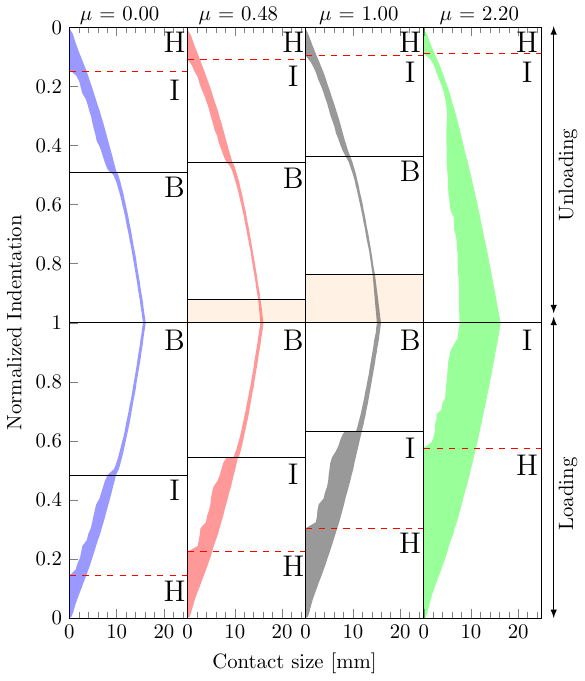}
    \end{subfigure}
 \caption[justification=justified]{The evolution of contact size and location in the loading and unloading phases for different CoF. The indentation on the y axes is normalized by the maximum indentation. The contact size represents the radius of a circular contact in the Hertzian regime and the width of an annular contact area in the intermediate and post-buckling regimes. Letters H, I, and B denote Hertzian, intermediate, and post-buckling regimes respectively.  Red dash lines mark the transition point between the Hertzian and intermediate regimes. The shaded areas highlight the elastic unloading at the onset of unloading for cases of $\mu = 0.48$ and $1.0$.\justifying 
}
    \label{fig:Fig4}
\end{figure}

Another intriguing effect of friction is that it makes the loading-unloading response asymmetric. Fig.~\ref{fig:Fig4} demonstrates the evolution of contact morphology during the loading and unloading phases by showing the location and size of the contact area. The contact size refers to the radius of a circular contact in the Hertzian regime and the width of the annular contact area in the intermediate and post-buckling regimes. Noting the variation of contact size, three regimes of Hertzian, intermediate, and post-buckling are classified for each friction case. It can be seen that friction postpones all transition points during the loading and unloading phases, resulting in a bigger Hertzian regime during the loading, but a smaller one upon unloading. Additionally, the degree of asymmetry in the loading-unloading response increases with friction. For the high friction case ($\mu = 2.2$) where the buckling regime is absent, unloading is mainly accommodated via the rolling mechanism, explaining the absence of dissipation (also see Supp. Movie~\cite{SuppInf}). 


We have shown how the interplay between structural instability and the local friction coefficient dictates the morphology of contact. It is shown that the contact behavior departs from the Hertzian solution due to bending at the onset of contact. This observation challenges the previous assumption of a Hertzian-like response in the pre-buckling regime. In agreement with previous studies ~\cite{audoly2010elasticity}, two transition points are identified in the loading phase, classifying three regimes of deformation: Hertzian, intermediate, and post-buckling. While the contact area increases monotonically in the first two regimes, it abruptly drops and then remains constant in the post-buckling regime. The distribution of pressure in the contact zone revealed two distinct mechanisms for the evolution of the contact area: (i) the energy-preserving rolling mechanism in the Hertzian and intermediate regimes, and (ii) the dissipative frictional sliding in the post-buckling regime. In other words, the response is fully reversible (non-dissipative) in the absence of friction or the post-buckling regime. The unloading phase for $\mu=0.48$ begins with an elastic unrolling, followed by dissipative frictional sliding. Furthermore, it is evident that friction causes a delay at all transition points, leading to the expansion and contraction of the Hertzian regime during the loading and unloading phases, respectively.

The interplay between friction, material properties, and structural instability leverages the controlled departure from Hertzian behavior and monotonic contact force-area relation. This new understanding presents a direction to tailor the contact morphology and governing mechanism, offering solutions to a diverse range of engineering applications such as systems with superior damping capabilities~\cite{ullah2022review}, sport ball~\cite{remond2022dynamical,rinaldi2016non}, switch haptic~\cite{plotnick2017force,gedsun2022bending}, or vesicle manipulation~\cite{quemeneur2012gel,tsapis2005onset}.\newline


%


\textbf{Acknowledgments.} This research has been supported by a Marie Sklodowska-Curie Postdoctoral Fellowship under contract number (proposal number 101065669). R.A. and R.S. also acknowledge the support from the Thomas B. Thriges Fond. M.A.D. would like to thank UKRI for support under the EPSRC Open Fellowship scheme (Project No. EP/W019450/1).

\newpage
\bibliography{apssamp}
\clearpage

\end{document}


\title{Supplementary information: Frictional contact of soft polymeric shells
\label{SI}}%

\author{Riad Sahli~\orcidlink{0000-0001-9077-3602}}%
\affiliation{%
Department of Mechanical and Production Engineering, Aarhus University, Denmark}%
\author{Jeppe Mikkelsen}%
\affiliation{Department of Mechanical and Production Engineering, Aarhus University, Denmark}%
\author{Mathias Sätherström Boye}%
\affiliation{Department of Mechanical and Production Engineering, Aarhus University, Denmark}%
\author{Marcelo A. Dias~\orcidlink{0000-0002-1668-0501}}
\email[Corresponding Author: ]{Marcelo.Dias@ed.ac.uk}
\affiliation{Institute for Infrastructure and Environment, School of Engineering, The University of Edinburgh, EH9 3FG Edinburgh, UK}
\author{Ramin Aghababaei~\orcidlink{0000-0002-0700-0084}}%
\email[Corresponding Author: ]{ra@mpe.au.dk}
\affiliation{%
Department of Mechanical and Production Engineering, Aarhus University, Denmark}%

\maketitle

In this supplementary information, we provide additional details, figures, and calculations that support our main findings.\newline

\textbf{Indentation setup.} We used a tensile machine (ZwickRoell) with a load cell (ZwickRoell Xforce HP 500N) to apply uniaxial load with an indentation velocity of 10 $mm.min^{-1}$. The hemispherical shell is attached to the loading cell at the top. A camera (Canon EOS 750D) with a lens (Canon macro EF-S 35mm F1:2.8 IS STM) is placed on the side of the sample. The image resolution is 1280×720 pixels and the acquisition rate is 50 frames per second. The system is illuminated by a diffused light on the opposite side of the camera (Powerfix FCL-80115 2400 lm 4000 K light source). This allows us to measure the normal load and the lateral profile at the same time.
\begin{figure}[h]
    \begin{subfigure}[b]{\linewidth}
         \centering
         \includegraphics[width=\textwidth]{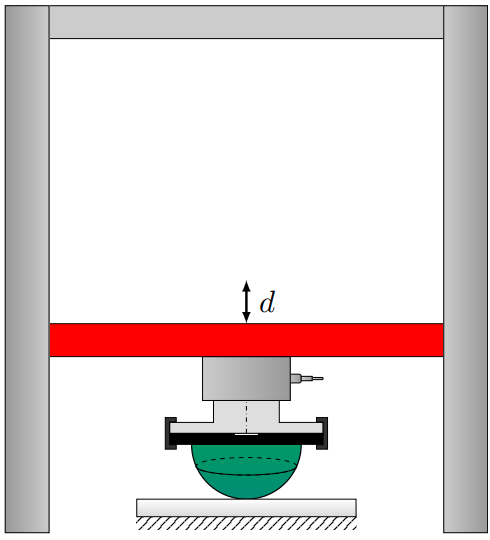}
    \end{subfigure}
 \caption[justification=justified]{The experimental setup sketch of a commercial ZwickRoell testing machine, with a nominal load of 5 kN and a ZwickRoell Xforce HP 500 N load cell. Shell in green is pressed on the substrate with imposed displacement $d$.
\justifying 
}
    \label{figSI:fullsetup}
\end{figure}
 For sample preparation, we used cross-linked polydimethylsiloxane (PDMS) elastomer base/curing agent mixture (mass ratio 10:1) of Sylgard 184, Dow Corning~\cite{le2009comparison}. The smooth hemispherical PDMS shells were obtained by pouring the liquid on top of a steel ball of radius 25 mm (supplier WILHELMSEN A/S, steel ball 50 mm, ref. KU 50G500) as described in~\cite{lee2016fabrication}. Thickness was measured postmortem using a profilometer (Keyence VR-3000 series). The shells are then fixed to a sample holder using PDMS. A hole in the sample holder allows air to freely escape, maintaining the atmospheric pressure during the experiment. The counterpart is composed of a polylactic acid (PLA) plane substrate with a roughness $Sa = 15$ $\mu m$ (ISO 25178-2:2012).\newline
 
\begin{figure}[b]
    \begin{subfigure}[b]{\linewidth}
         \centering
         \includegraphics[width=\textwidth]{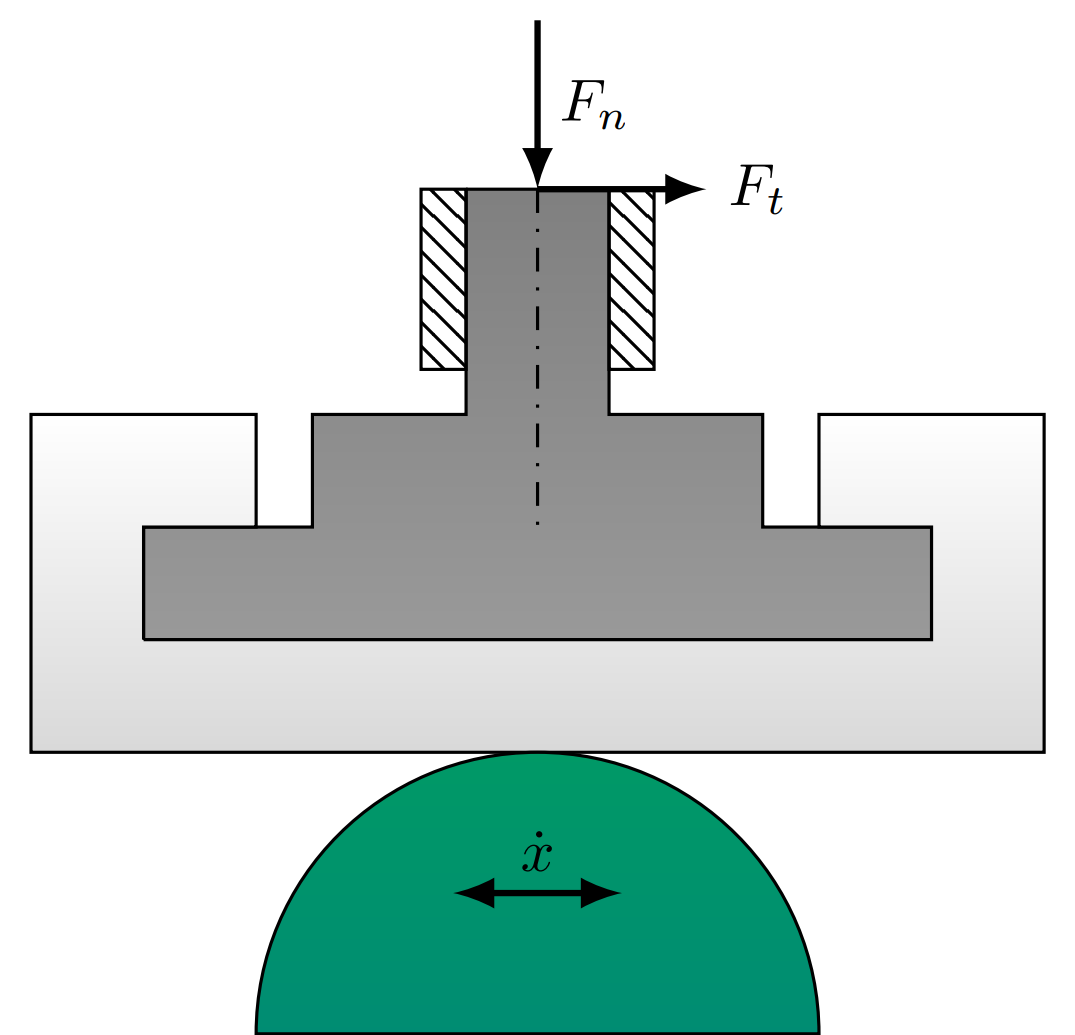}
    \end{subfigure}
 \caption[justification=justified]{The setup for measuring CoF between the shell and rigid plate. The transparent section, denoted as the test object, is pressed into the underlying green shell by adjusting the normal force ($F_n$). Subsequently, the underlying material is oscillated in a back-and-forth motion at a sliding velocity ($\dot x$), during which the tangential force ($F_t$) is extracted.\justifying 
}
    \label{figSI:friction}
\end{figure}

\textbf{Friction setup} The friction coefficient ($\mu$) for PDMS on PLA substrate used in simulations is obtained using additional friction experiments using a tribometer (Rtec-instruments SMT-5000). The system is composed of a plain PDMS hemisphere with a radius of 15 mm, put in contact with a rigid plate of the PLA substrate (Fig.~\ref{figSI:friction}). We used a range of normal forces [0.25, 5] N, displacement of 20 mm, and velocities of $1.2$, $1.4$, $1.6$, and $1.8$ $mm.s^{-1}$. Seven tests are performed for each velocity and the average friction coefficient is calculated. To study the effect of friction, a lubricant (Kema GL-68) is applied to the system. The following CoFs are obtained for the dry PDMS/PLA $\mu= 1.0  \pm 0.2$ and Lubricated PDMS/PLA $\mu=0.48 \pm 0.09$ contacts. \newline

\begin{figure}[t]
    \begin{subfigure}[b]{\linewidth}
         \centering
         \includegraphics[width=\textwidth]{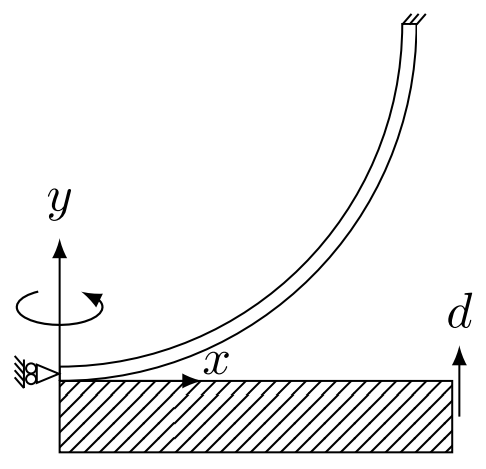}
    \end{subfigure}
 \caption[justification=justified]{The axisymmetric hemispherical shell is modeled to be in contact with an analytically rigid plane (hatched). The hemispherical shell is clamped on the top surface and displacement is applied to the rigid plane in y-direction.\justifying 
}
    \label{figSI:abacus}
\end{figure}

\textbf{Numerical setup} We used the commercial software Abaqus to conduct finite element simulations. The simulations are used to extract the contact area and corresponding force. The simulation setup contains an axisymmetric model, consisting of a deformable shell and a rigid plate. The gravitational effect is accounted for. CAX3H elements and contact penalization are used. Normal reaction force, global area, local contact area, pressure, and deformation are extracted from the simulation. The top of the sphere is locked in the y-direction, while the indentation is applied by moving the rigid plate upwards in the y-direction (Fig.~\ref{figSI:abacus}). Friction is added using the penalty method based on the Coulomb friction law, which relates the friction force to the contact pressure. Abaqus/Explicit is used as the solver, utilizing automatic stabilization. The stabilization parameters are reduced systematically to ensure that the solution is not disturbed, with the final value being a dissipated energy fraction of 0.1\textperthousand{} with a maximum stabilization to strain energy ratio of 0.5\textperthousand{}. The material and geometrical parameters used in the simulations are:
Density of polymer = 1100,
Sphere radius = 25 mm,             
Youngs modulus polymer = 1.6 MPa,
Poissons ratio polymer = 0.45 and
Mesh size = 800.

\begin{figure}[t]
     \centering
        \includegraphics[width=\linewidth]{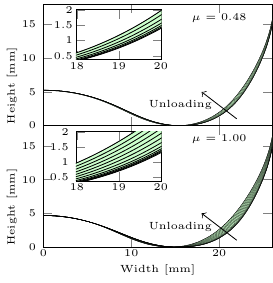}
        \caption[justification=justified]{
         The shell profile at the onset of the unloading phase for $\mu = 0.48$ and $\mu = 1.00$ obtained from simulations. As seen, the buckled (inner) side of the shell remains fairly stationary while the outer side unrolls. \justifying }
    \label{figSI:Fig2}
\end{figure}
\begin{figure}[t]
    \begin{subfigure}[b]{\linewidth}
         \centering
         \includegraphics[width=\textwidth]{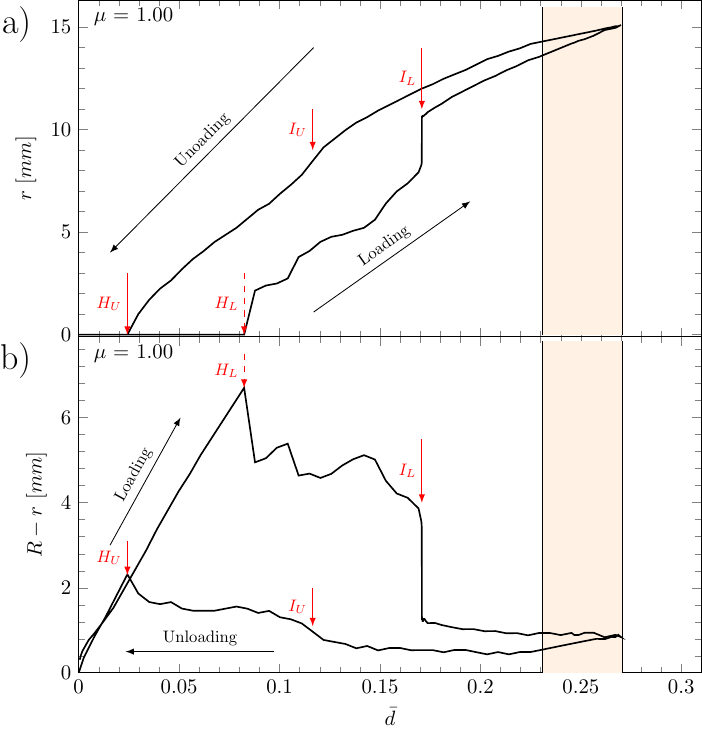}
    \end{subfigure}
 \caption[justification=justified]{(a) The evolution of the inner contact radius $r$ and (b) the contact width as a difference between the inner and outer radius $R-r$ for $\mu = 1.00$ . $H_L$ and $H_U$ denote the departure from and to the Hertzian solution in the loading and unloading phases (due to bending/unbending). $I_L$ and $I_U$ denote the point of instability and shell buckling/unbuckling in the loading and unloading phases. The shaded area marks the stage where the outer side of the shell unloaded elastically.\justifying 
}
    \label{figSI:radius}
\end{figure}

Fig. \ref{figSI:Fig2} shows the onset of unloading, where the outer side is unloaded elastically first. In this regime, while the buckled (inner) side of the shell remains fairly stationary due to friction, the outer side unrolls, causing a reduction in the contact area. The inner $r$ and the outer $R$ contact radius provide precious information about the evolution of contact area at different stages of loading and unloading. During the post-buckling phase ($p_2$ and $p_3$), the buckled (inner) side of the shell remains relatively stationary due to friction (see Fig.~\ref{figSI:radius}a), while the outer side unrolls, leading to a decrease in the contact area (Fig.~\ref{figSI:radius}b).

Table \ref{tab:1} reports the energy dissipation upon unloading $W$ and the drop in the contact area $\Delta A$ at the point of buckling as a function of CoF for both experiment and simulation. Values are extracted from the data in Fig.~\ref{fig:Fig1}a and \ref{fig:Fig1}b respectively.

\begin{table}[h]
\centering
\begin{tabular}{|c  c  c|}
\hline
  & $\Delta A$ [$mm^2$]& $W$ [$mJ$] \\
  \hline
  \hline

 $\mu_{0.48}$ $ exp.$ & - & 2.2 $\pm$ 0.1 \\ 
 $\mu_{1.00}$ $ exp.$ & - & 3.70 $\pm$ 0.04 \\ 
    \hline
   $\mu_{0.00}$ $ sim.$ & 63 & 0.00 \\  
 $\mu_{0.48}$ $ sim.$ & 99 & 1.80 \\  

 $\mu_{1.00}$ $ sim.$ & 162 & 3.05 \\  
 $\mu_{2.20}$ $ sim.$ & - & 0.44 \\ 
\hline
 $\mu_{1.50}$ $ extra$ $sim.$ & 208 & 3.11 \\  
 $\mu_{2.00}$ $ extra$ $ sim.$ & 263 & 2.04 \\
 
 \hline
\end{tabular}
 \caption[justification=justified]{
Energy dissipation upon unloading $W$ and the drop of contact area at the point of buckling $\Delta A$, obtained from Fig.~\ref{fig:Fig1}a and b. \justifying
}
    \label{tab:1}
\end{table}

\newpage
\bibliography{apssamp}